\documentclass[aps,twocolumn,secnumarabic,balancelastpage,amsmath,amssymb,nofootinbib]{revtex4}
\usepackage{natbib}
\usepackage{color}           
\usepackage{graphics}         
\usepackage[pdftex]{graphicx}  
\usepackage{longtable}        
\usepackage{epsfig}
\usepackage{bm}             
\usepackage{thumbpdf}

\usepackage[colorlinks=true]{hyperref}  

\begin{document}

\title{Effects of disorder on carrier transport in Cu$_2$SnS$_3$}
\author{Lauryn L. Baranowski$^{1,2}$}
\author{Kevin McLaughlin$^2$}
\author{Pawel Zawadzki$^1$}
\author{Stephan Lany$^1$}
\author{Andrew Norman$^1$}
\author{Hannes Hempel$^3$}
\author{Rainer Eichberger$^3$}
\author{Thomas Unold$^3$}
\author{Eric S. Toberer$^{1,2}$}
\author{Andriy Zakutayev$^1$}
\affiliation{$^1$National Renewable Energy Laboratory, 15013 Denver West Parkway, Golden CO 80401, USA}
\affiliation{$^2$Physics Department, Colorado School of Mines, 1500 Illinois St., Golden CO 80401, USA}
\affiliation{$^3$Helmholtz Center Berlin for Materials and Energy, Hahn-Meitner-Platz 1, D-14109 Berlin, Germany}

\begin{abstract}

In recent years, further improvements in the efficiency of Cu$_2$ZnSn(S,Se)$_4$ photovoltaic devices have been hampered due to several materials issues, including cation disorder. Cu$_2$SnS$_3$ is a promising new absorber material that has attracted significant interest in recent years. However, similar to CZTS, Cu$_2$SnS$_3$ displays cation disorder. In this work, we develop synthetic techniques to control the disorder in Cu$_2$SnS$_3$ thin films. By manipulating the disorder in this material, we observe crystal structure changes and detect improvements in the majority carrier (hole) transport. However, when the minority carrier (electron) transport was investigated using optical pump terahertz probe spectroscopy, minimal differences were observed between the ordered and disordered Cu$_2$SnS$_3$. By combining these results with first-principles and Monte Carlo theoretical calculations, we are able to conclude that even ostensibly ``ordered'' Cu$_2$SnS$_3$ displays minority carrier transport properties corresponding to the disordered structure. The presence of extended planar defects in all samples, observed in TEM imaging, suggests that disorder is present even when it is not detectable using traditional structural characterization methods. The results of this study highlight some of the challenges to the further improvement of Cu$_2$SnS$_3$-based photovoltaics, and have implications for other disordered multinary semiconductors such as CZTS.

\end{abstract}

\maketitle

\noindent {\textbf{Introduction}}

The development of efficient and scalable photovoltaics is an important challenge in today's energy landscape. Although highly efficient thin film photovoltaics have been commercially available for some time, these technologies may be limited in their scalability by the use of rare or toxic elements \cite{Wolden2011,Woodhouse2013,Kavlak2015}. Cu$_2$SnS$_3$ is a promising thin film photovoltaic absorber material which uses only abundant and nontoxic elements. Interest in this compound has rapidly increased in the past several years, and device efficiencies of 4\% and 6\% have been reported using Cu$_2$SnS$_3$-based absorber materials and alloys, respectively \cite{Kanai:2014vu,Maekawa:2013fw}.

Besides the reported device efficiencies, previous theoretical and experimental works have revealed promising attributes of Cu$_2$SnS$_3$ for photovoltaic absorber applications. First principles calculations showed a wide range of phase stability, high optical absorption coefficient, and lack of Fermi level pinning in this material \cite{Zawadzki:2013fr}. Experimental reports of the band gap of Cu$_2$SnS$_3$ range from 0.9-1.35 eV, well suited for a single-junction PV device \cite{Berg:2012it,Fernandes:2010fc}. Furthermore, alloying with Si or Ge on the Sn site can serve to increase the band gap. Research has demonstrated that Cu$_2$SnS$_3$ can be synthesized by several potentially scalable techniques such as sputtering and solution processing \cite{Koike2012,Avellaneda2010,Maekawa:2013fw}.

A variety of crystal structures have been reported for Cu$_2$SnS$_3$ thin films, including cubic, tetragonal, monoclinic, and triclinic \cite{Avellaneda2010,Maekawa:2013fw,Chino2012}. First principles calculations for this compound suggest that the different crystal structures result from Cu/Sn disorder on the cation sites of the zinc blende-based lattice. In Ref. \cite{Zhai:2011fc}, the authors calculate the structures resulting from varying degrees of random disorder, ranging from cubic (fully disordered), to tetragonal (partially disordered), to monoclinic (fully ordered). A recent theoretical work has suggested that this disorder takes the the form of compositional inhomogeneities caused by entropy-driven clustering (rather than fully random cation disorder), and could lead to potential fluctuations that negatively affect the carrier transport in Cu$_2$SnS$_3$ \cite{Zawadzki2015}. It has been demonstrated that annealing Cu$_2$SnS$_3$ thin films at higher temperatures promotes a transformation from the tetragonal to the monoclinic structure \cite{Chalapathi:2013dh}. A few other studies have considered annealing of Cu$_2$SnS$_3$, including an investigation of the behavior of extrinsic oxygen defects in this material \cite{Tiwari:2014bl}. However, there has been no work on the behavior of intrinsic defects during the annealing process, or the effects of annealing on the electronic properties of Cu$_2$SnS$_3$.

The reported hole concentrations in Cu$_2$SnS$_3$ are often high, ranging from 10$^{17}$ -10$^{19}$ cm$^{-3}$ \cite{Aihara:2013ft,Su:2012ja}. This level of doping is generally considered too high for a photovoltaic absorber material, as it leads to a tunneling-enhanced increase in recombination within the absorber layer and at the heterojunction interface. In our prior work, we established that both Cu-poor and S-poor conditions are necessary for low carrier concentrations \cite{Baranowski:2014em}. By controlling both the Cu and S chemical potentials during film growth, we were able to tune the carrier concentration over 3 orders of magnitude and achieve films with p-type doping of less than 10$^{18}$ cm$^{-3}$. Even with this progress, lower carrier concentrations are still desirable for device integration, and have proven difficult to achieve through control of the Cu and S chemical potentials during film growth. When the Cu-poor and S-poor requirements are considered in conjuction with the Cu-Sn-S chemical potential phase space \cite{Zawadzki:2013fr}, it can be determined that the lowest carrier concentrations should be found in Cu$_2$SnS$_3$ that is in equilibrium with SnS.

In this work, we equilibrated our as-deposited Cu$_2$SnS$_3$ with SnS by annealing the films under an SnS atmosphere. We found that this equilibration with SnS caused a transformation from the cubic/tetragonal structure to the monoclinic structure, along with a decrease in the hole concentration. We used optical pump terahertz probe spectroscopy to investigate the minority carrier transport in our Cu$_2$SnS$_3$ films. Both samples had short electron decay times of 0.1-10 ps, and similarly high degrees of charge localization, characteristic of a disordered structure. When the samples were imaged with transmission electron microscopy, both the cubic and monoclinic films had high densities of stacking faults and/or twins. We conclude that the existence of planar defects results in local disorder even in the monoclinic Cu$_2$SnS$_3$, and has a negative impact on the electronic transport in this material. The findings in this work have implications not only for Cu$_2$SnS$_3$, but also for Cu$_2$ZnSn(S,Se)$_4$ (CZTS) and other multinary disordered semiconductors. \\

\noindent {\textbf{Methods}} \\

\noindent {\it{Film synthesis and characterization}} 

The films described in this paper were deposited using combinatorial RF sputtering from 50 mm diameter Cu$_2$S and SnS$_2$ targets on heated 50x50 mm glass substrates, under conditions described in Ref. \cite{Baranowski:2014em}. Further information about combinatorial synthesis approaches as applied to other absorber materials can be found in Ref.s \cite{Green:2013iy,Zakutayev2015,Welch2015,Caskey2014}. Characterization was performed on small (12.5 mm x 12.5 mm) sections of the original films, which were determined to be uniform with regards to composition and morphology (due to the shallow initial compositional gradient and size of film sections). X-ray fluorescence spectra were collected using a Fischer XDV-SDD instrument to obtain both the Cu/Sn ratios and the thickness of the films. The X-ray diffraction patterns were collected using a $\theta - 2 \theta$ geometry with Cu K-$\alpha$ radiation and a proportional 2D detector (Bruker D8 Discover with General Area Detector Diffraction System software). Raman spectra were collected using a Renishaw inVia confocal Raman microscope configured with 532 nm laser excitation at 5\% power, an 1800 mm$^{-1}$ grating and CCD array detector. To compensate for the small spot size of the Raman measurement ($\sim$10 $\mu$m), six measurements were taken on each sample and averaged to provide a final spectrum. 


We used optical pump terahertz probe spectroscopy (OPTP), also known as time resolved terahertz spectroscopy (TRTS), to measure the decay of pump induced conductivity in the ps range as well as  the complex charge carrier mobility. The OPTP spectrometer has been described earlier \cite{Stroth2012}. The setup has been modified recently to record the pump induced change in THz reflection configuration. The transients were measured at the point of maximum change in THz amplitude, which represents the averaged THz conductivity decay, while the mobilities were analyzed by fitting the recorded THz reflection spectra  to an optical model using the transfer matrix method. We assumed that the decay in conductivity is caused by a decaying carrier concentration where the maximum of the transient corresponds to all initially induced carriers. While the measured mobility is an average over all excited charge carriers, the conductivity is dominated by the electron (minority carrier) properties if the electron mobility can be assumed to be larger than the hole mobility \cite{Stroth2014}.  \\

\noindent {\it{Annealing experiments}} 

SnS powder was synthesized using Sn (Alfa Aesar, 99.9\%) and S (Alfa Aesar, 99.5+\%) powders in a stoichiometric ratio. The reactants were sealed under static vacuum ($<$5 mTorr) in a quartz ampoule, heated to 400$^{\circ}$C, and held for 10 hrs. Then, the reactancts were heated to 900$^{\circ}$C and held for 18 hrs, before cooling to room temperature. This resulted in a phase pure SnS powder. The SnS$_2$ powder was synthesized by ball-milling the SnS powder with excess S for 2 hrs in a SPEX 8000D Mixer/Mill. The ball milled mixture was sealed in a quartz ampoule under static vacuum and held at 425$^{\circ}$C for 18-24 hrs, resulting in a phase pure SnS$_2$ powder.

Annealing was performed by placing the samples in a quartz tube along with $\sim$1 g of SnS or SnS$_2$ powder in an alumina boat. A thermocouple was inserted into the tube such that the thermocouple was positioned directly over the sample. Flowing argon gas was supplied to the quartz tube, and exited via a bubbler filled with mineral oil. The samples were heated in a tube furnace with a 5 hour total ramp time to the desired temperature. After the desired hold time, the furnace was turned off and allowed to cool to room temperature. \\

\noindent {\it{Computational methods}} 

Density of states (DOS) and projected density of states were calculated using the VASP code \cite{Kresse1999}. Because of the band gap problem of the semi-local density functional
theory (DFT) and large cells necessary to describe disorder in Cu$_2$SnS$_3$ we performed non-self consistent HSE calculations \cite{Heyd2003}. The wave functions for HSE were generated with DFT+$U$ \cite{Dudarev1998} using Perde-Burke-Ernzerhof \cite{Perdew1996,Perdew1997} exchange correlation functional and on-site potential $U=7$ eV applied to Cu($d$) states. 

To compare degree of charge localization in the ordered (SG=$Cc$) and disordered Cu$_2$SnS$_3$ we also calculated inverse participation ratio (IPR). IPR measures the inverse of the fraction of atoms over which a given state is delocalized and takes the form
\begin{equation}
IPR = \frac{N \sum_i^N c_i^4}{(\sum_i^N c_i^2 )^2 }
\end{equation}
where the sums run over $N$ atoms in the unit cell and $c_i$ are atom-projected density of states. For instance, for a state that is delocalized over all atoms in the unit cell, $IPR=1$; for a state delocalized over half of the atoms in the unit cell, $IPR=2$.

For the disordered Cu$_2$SnS$_3$, the DOS and IPR were averaged over four independent atomic structures generated using Metropolis Monte Carlo method with a local motif-based model Hamiltonian described in Ref. \cite{Zawadzki2015}. \\

\noindent {\textbf{Results and discussion}} \\

\begin{figure}
   		 \includegraphics[width=7cm]{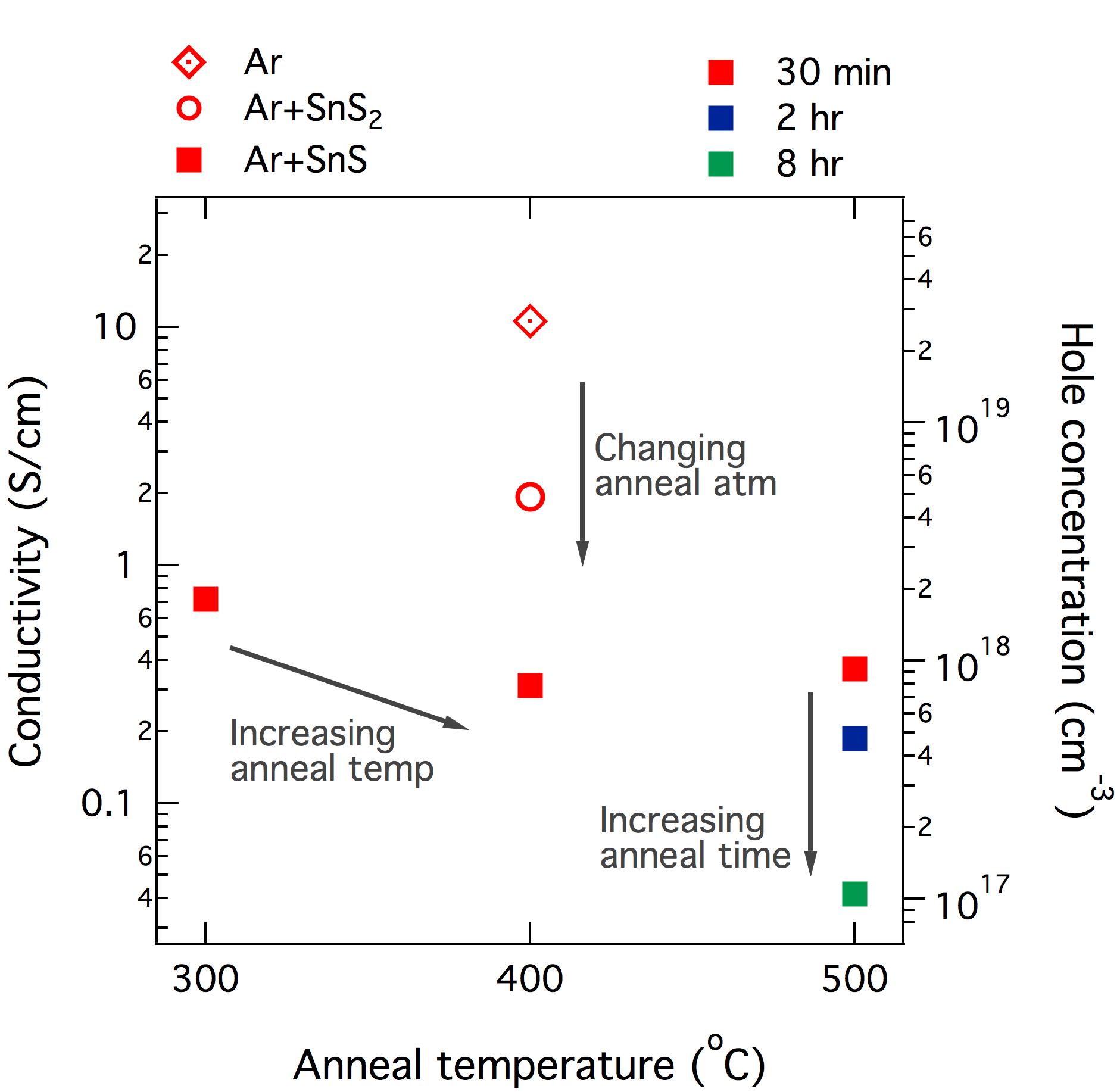} 
\caption{As expected from the Cu-Sn-S chemical potential phase space, carrier concentration reductions are largest when the Cu$_2$SnS$_3$ films are annealed in equilibrium with SnS (square markers). Increasing anneal temperature shows some effect on the hole concentration; increasing anneal time causes more dramatic reductions.}
\label{Fig:atm}
\end{figure}

We investigated a variety of annealing conditions to determine their effects on the transport properties of Cu$_2$SnS$_3$. The annealing temperature, atmosphere, and duration were independently explored to determine the effects on the carrier concentration, and these results are presented in Fig. \ref{Fig:atm}. As expected from the chemical potential phase space, the lowest hole concentrations were achieved under an SnS/Ar atmosphere (square data points in Fig. \ref{Fig:atm}) \cite{Baranowski:2014em}. Comparing the results for 30 min, 2 hrs, and 8 hrs (red, blue, and green data points, respectively), we found that the kinetics of the annealing process were relatively slow. We determined that hole concentrations of $\sim$5$\cdot$10$^{17}$ cm$^{-3}$ could be achieved by annealing under an SnS/argon atmosphere at 500$^{\circ}$ for 2 hours. These anneal parameters were used to investigate the structural and morphological changes that resulted from annealing, as well as both the majority (holes) and minority (electrons) carrier transport. Although lower carrier concentrations were achieved by annealing for 8 hrs (1$\cdot$10$^{17}$ cm$^{-3}$), the length of the anneal would preclude this sample from any kind of industrial scale-up, and as such, we have excluded it from further investigation. \\

\noindent {\textit{Structural changes \& majority carrier transport}}

We assessed changes in the crystal structure of the Cu$_2$SnS$_3$ films using X-ray diffraction (XRD) and Raman spectroscopy. Overall, we saw a change from a cubic/tetragonal as-deposited structure, to a monoclinic post-annealed structure. This indicated a change from a disordered cation sublattice (cubic/tetragonal) to an ordered sublattice (monoclinic). 

\begin{figure}
   		 \includegraphics[width=7cm]{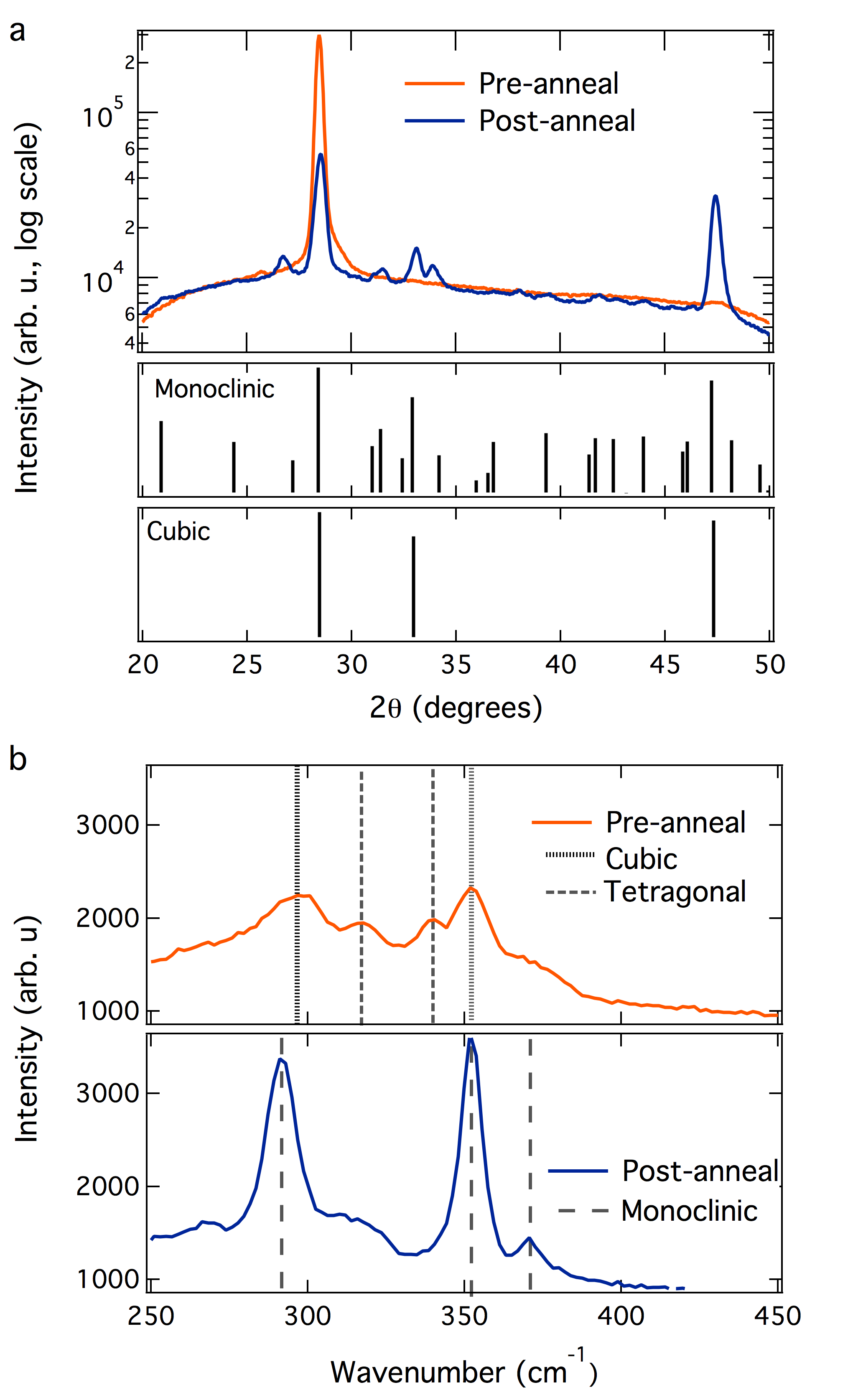} 
\caption{(a) After annealing, the Cu$_2$SnS$_3$ films display XRD patterns corresponding to the monolinic structure. Prior to annealing, structural determination is difficult using only one peak. (b) Raman spectroscopy allows for elucidation of the pre-anneal structure as a mixture of cubic and tetragonal Cu$_2$SnS$_3$. The post-annealed spectrum shows peaks corresponding the monoclinic structure, as expected from the XRD pattern. }
\label{Fig:ordering}
\end{figure}

Prior to annealing, the XRD pattern displayed only one peak at $\sim$28.5$^{\circ}$, shown in Fig. \ref{Fig:ordering}a. With only this peak, it is difficult to assign a definitive crystal structure to this XRD pattern: this peak could correspond to the cubic, tetragonal, or monoclinic structures of Cu$_2$SnS$_3$. Thus, we used Raman spectroscopy to further probe the structure of the preferentially oriented films. The pre-anneal Raman spectrum, shown in Fig. \ref{Fig:ordering}b, showed two major peaks at 299 and 351 cm$^{-1}$, corresponding to the cubic Cu$_2$SnS$_3$ structure. Secondary peaks at 317 and 338 cm$^{-1}$ could be ascribed to the tetragonal crystal structure \cite{Fernandes:2010fc}. 

Post-anneal, the XRD pattern showed peaks corresponding to the monoclinic Cu$_2$SnS$_3$ structure. The Raman spectrum had two major peaks at 290 and 351 cm$^{-1}$, which can be assigned to the monoclinic structure \cite{Berg:2012ks}. Two secondary peaks can be seen at 315 and 371 cm$^{-1}$. In Ref. \cite{Berg:2012ks} these peaks are ascribed to a Cu$_2$Sn$_3$S$_7$ secondary phase, with the corresponding cation ratio (determined by energy dispersive X-ray spectroscopy) of Cu/(Cu+Sn)=0.47. However, the cation ratio of our film, as determined by X-ray fluorescence, was Cu/(Cu+Sn)=0.63, suggesting that the amounts of secondary phase present are minimal.

Concurrent with the change in crystal structure upon annealing, we observed an improvement in the majority carrier transport. The carrier concentration decreased from 1.3$\cdot$10$^{19}$ cm$^{-3}$ for the as-deposited sample to 8.0$\cdot$10$^{17}$ cm$^{-3}$ after annealing. Additionally, the Hall mobility increased from 0.56 cm$^2$/Vs to 8.2 cm$^2$/Vs. Two possible explanations for the increase in hole mobility are a reduction in grain boundary scattering due to grain growth during annealing, or a reduction in ionized defect density. Measuring the Hall mobility of the pre-annealed samples as a function of temperature shows the expected increase in mobility at low temperatures, suggesting that the hole transport is not grain boundary limited (see Fig. \ref{Fig:tdhall}). Thus, we conclude that the reduction in ionized defect density is responsible for the increased hole mobility. \\

\noindent {\textit{Terahertz spectroscopy investigation of minority carrier transport}}

In order to probe the minority carrier transport in the as-deposited and annealed Cu$_2$SnS$_3$, we performed optical pump terahertz probe spectroscopy (OPTP). In contrast to the improvement observed in the hole (majority carrier) transport, the changes in electron (minority carrier) transport after annealing were less pronounced. 

\begin{figure}
   		 \includegraphics[width=7cm]{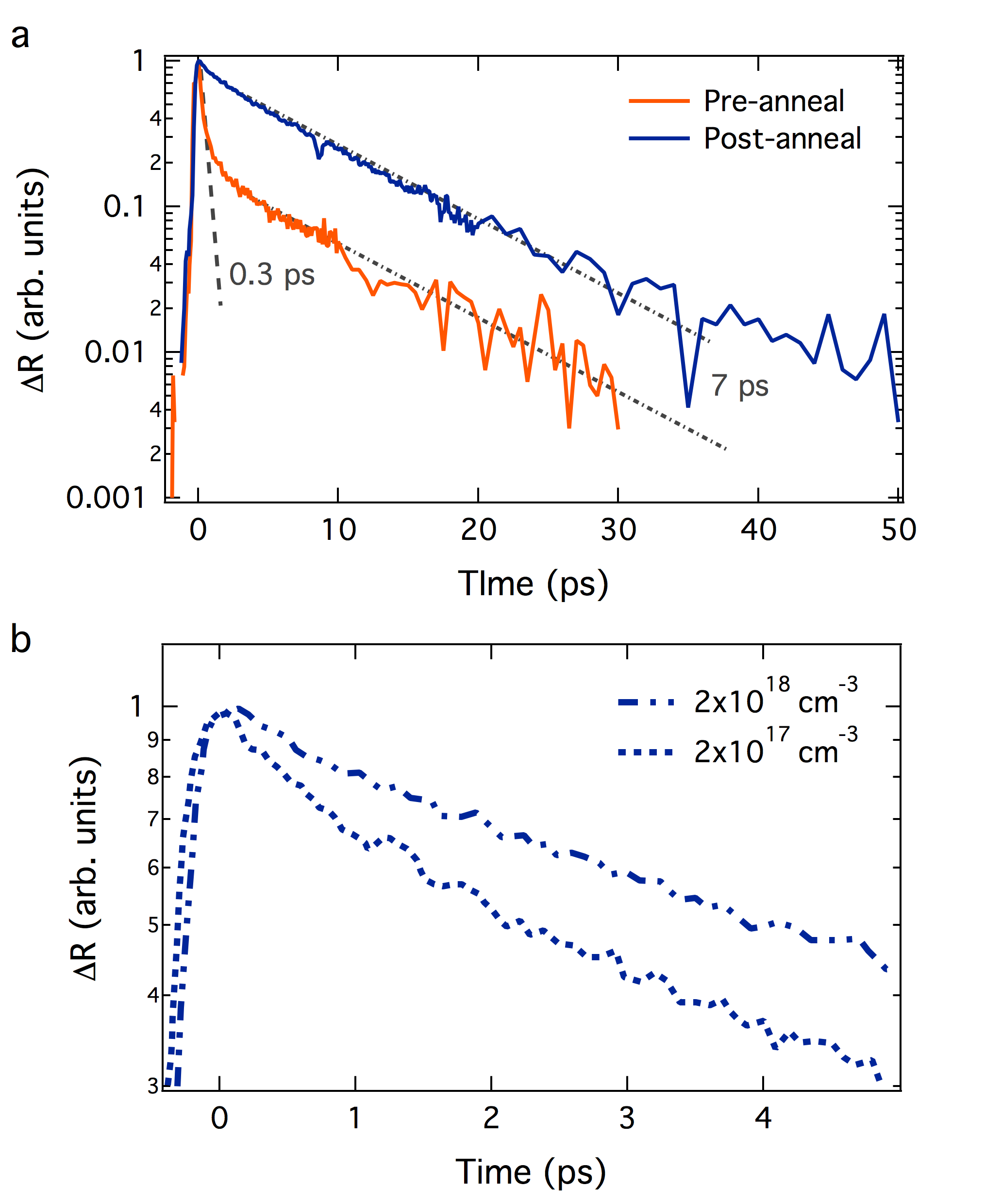} 
\caption{(a) The as-deposited sample shows two decay processes in the measured reflectivity: a 0.3 ps decay and a 7 ps decay; the annealed sample shows only the 7 ps decay process. (b) When the excitation power is lowered, the 0.3 ps decay can be detected in the annealed sample, suggesting that this state can be saturated for high carrier injection.}
\label{Fig:Thz}
\end{figure}

In the as-deposited sample, the measured reflectivity as a function of time showed two decay processes: one with a decay time of 0.3 ps, and a second with a decay time of 7 ps (see Fig. \ref{Fig:Thz}a). When the reflectivity of the annealed Cu$_2$SnS$_3$ is compared to that of the as-deposited sample, it is clear that the 0.3 ps decay process is absent, but that the 7 ps decay process remains. However, when the excitation power is lowered to 2$\cdot$10$^{17}$ cm$^{-3}$ (as measured in terms of excited carriers at the surface of the sample), the 0.3 ps decay process can be detected even in the annealed sample, as shown in Fig. \ref{Fig:Thz}b. This suggests that the trap or defect states responsible for the 0.3 ps decay can be saturated for high carrier injection. At one sun conditions (lower than the excitation powers used here), this decay process may still be dominant.

\begin{table*}
\begin{center}
\begin{tabular}{| l | p{3.5cm} | p{3.5cm} |} 
\hline
  & Pre-Anneal & Post-Anneal \\ \hline
\multicolumn{3}{| l |}{{\bf{Structure}}} \\ \hline
Crystal structure (XRD) & Cubic, oriented & Monoclinic \\ \hline
Crystal structure (Raman) & Cubic/tetragonal & Monoclinic \\ \hline
Crystal structure (TED) & Cubic, oriented & Monoclinic \\ \hline
\multicolumn{3}{| l |}{{\bf{Morphology}}} \\ \hline
Grain shape & Columnar & Equiaxed \\ \hline
Grain size & $\sim$50 nm & 200-500 nm \\ \hline
Defects & High density of \{111\} planar defects & Some planar \& other intragrain defects \\ \hline
\multicolumn{3}{| l |}{{\bf{Majority carrier transport}}} \\ \hline
Hole concentration & 1$\cdot$10$^{19}$ cm$^{-3}$ & 8$\cdot$10$^{17}$ cm$^{-3}$ \\ \hline
Hole mobility & 0.56 cm$^2$/Vs & 8.2 cm$^2$/Vs \\ \hline
\multicolumn{3}{| l |}{{\bf{Minority carrier transport}}} \\ \hline
Electron decay times & 0.3 ps, 7 ps & 7 ps \\ \hline
Momentum relaxation time & 33 fs & 69 fs \\ \hline
Electron mobility & 26 cm$^2$/Vs & 55 cm$^2$/Vs \\ \hline
Electron localization constant & -0.91 & -0.91 \\ \hline
\end{tabular}
\end{center}
\label{table1}
\caption{Summary of changes in crystal stucture, film morphology, and majority and minority carrier transport caused by annealing Cu$_2$SnS$_3$ thin films. Significant changes were evident in the structure and morphology, and a concurrent improvement was observed in the majority carrier transport. However, the minority carrier transport was mostly unaffected by the annealing. }
\end{table*}

For both the as-deposited and annealed samples, the imaginary component of the terahertz conductivity is negative (Fig. \ref{Fig:Thzcond}). This negative value indicates charge carrier localization, which can be modeled using the Drude-Smith fit for conductivity \cite{Smith2001}. Fitting our complex conductivity data results in a determination of the relaxation scattering time $\tau$ and a constant $c_1$, which represents the persistence of velocity after the first scattering event. The constant $c_1$ can vary from -1 to 0, with $c_1$=0 representing Drude conductivity and $c_1$=-1 representing complete carrier localization. For the as-deposited and annealed Cu$_2$SnS$_3$, the calculated value of $c_1$ is -0.91 (calculated 5-10 ps after excitation), indicating strong localization in both samples. The relaxation scattering times were 33 fs and 69 fs, for the as-deposited and annealed samples, respectively.

The minority carrier DC mobility of the samples can also be calculated from the Drude-Smith model (with $\omega$=0 for DC mobility), if the effective mass (m$^{\star}$) is known or approximated. Here, we approximate $m^{\star}$ as 0.2$m_e$, the value calculated for CZTS \cite{Liu2012}. We calculate the minority carrier mobility to be 26 cm$^2$/Vs for the as-deposited Cu$_2$SnS$_3$, and 55 cm$^2$/Vs for the annealed sample. This factor of two increase in the electron mobility is directly related to the increase in the relaxation scattering time $\tau$. As with the increase in hole mobility, this is likely due to a reduction in ionized defect density after annealing. \\

\noindent {\textit{Theoretical explanation of terahertz spectroscopy results}}

\begin{figure}
   		 \includegraphics[width=7cm]{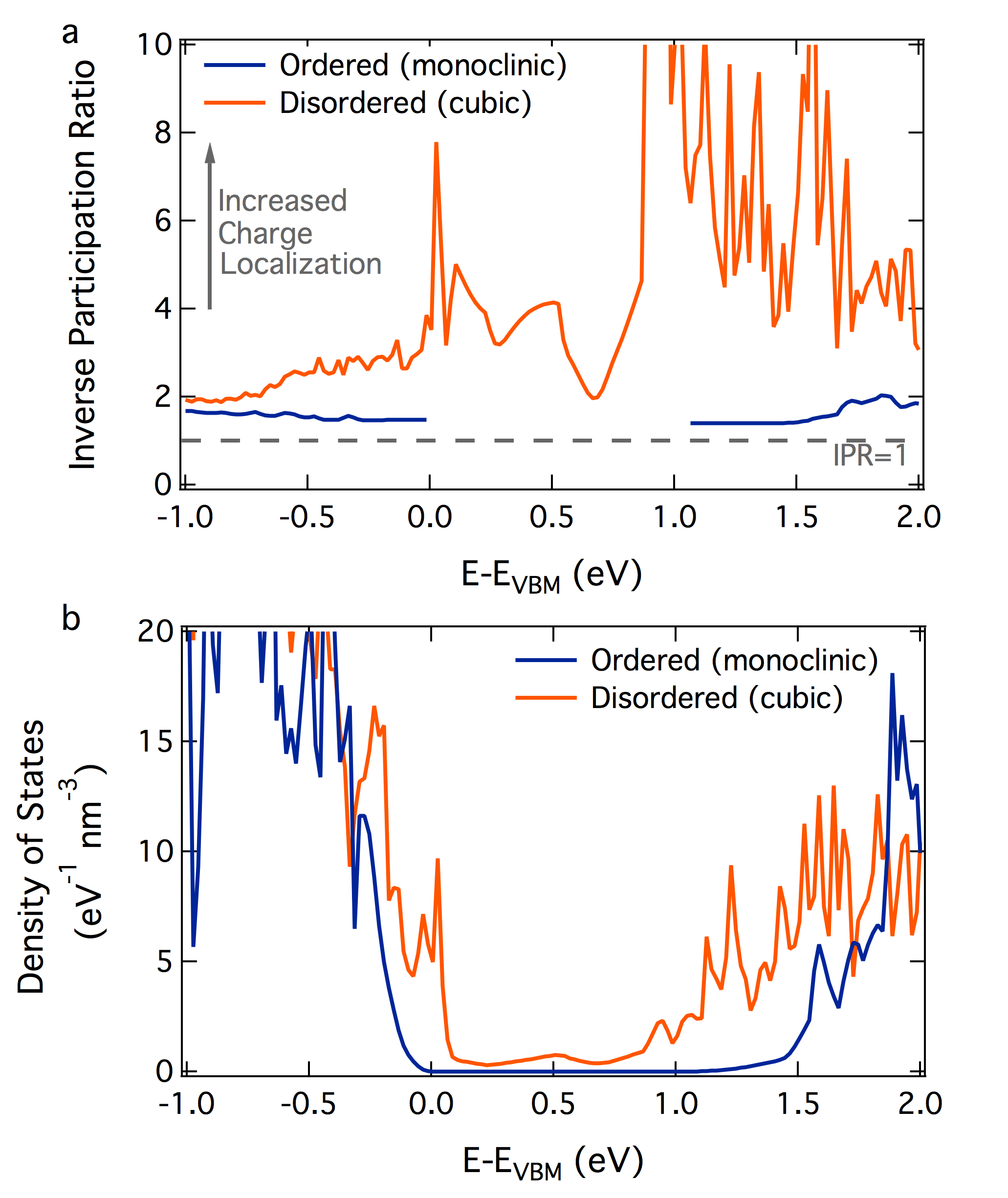} 

\caption{(a) Calcuation of the inverse participation ratio (IPR) for the disordered and ordered structures shows a high degree of charge localization (denoted by a high IPR value) for the disordered structure, suggesting that both of our samples exhibit disorder. (b) The disordered structure results in significant band tailing and a mid-gap state, both of which would significantly impact the performance of a Cu$_2$SnS$_3$-based photovoltaic device.}
\label{Fig:IPR}
\end{figure}

To interpret our finding that both the ordered and disordered samples displayed high degrees of charge localization, we turn to the calculated inverse participation ratios (IPR) for ordered and disordered Cu$_2$SnS$_3$. The IPR values represent the degree of charge localization: an IPR value of 1 indicates that charge is totally delocalized; an IPR value of 2 indicates that charge is localized on $\frac{1}{2}$ of the atoms present in the structure. As shown in Fig. \ref{Fig:IPR}a, the ordered Cu$_2$SnS$_3$ has IPR ratios varying between 1-2 at the band edges (similar to values for traditional semiconductors such as Si or GaAs). In contrast, the disordered structure has high IPR values, especially at the conduction band edge, indicating a high degree of charge localization. Thus, when charge localization is used as a metric, both of our experimental samples could be considered ``disordered'', as both display a high degree of charge localization ($c_1$). 

At this point in the interpretation of our results, it is necessary to summarize some of the recent important theoretical work regarding the nature of disorder in Cu$_2$SnS$_3$ \cite{Zawadzki2015}. The Cu$_2$SnS$_3$ structure can be understood as an assembly of S coordination motifs, in which each S anion is tetrahedrally coordinated by four cations. In the ground state, Cu$_2$SnS$_3$ is made up of only S-Cu$_2$Sn$_2$ and S-Cu$_3$Sn motifs; the S-Cu$_4$, S-CuSn$_3$, and S-Sn$_4$ motifs are higher energy motifs and are not present. In the monoclinic (ordered) structure, the S-Cu$_2$Sn$_2$ motifs are uniformly distributed throughout the supercell used in the computation. However, when the disordered structure is obtained from Monte Carlo simulations, the S-Cu$_2$Sn$_2$ motifs form nanometer-scale clusters within the supercell. This clustering is due to the presence of attractive entropic forces in the disordered structure. Thus, when we speak of disorder in Cu$_2$SnS$_3$, we refer to this clustering effect, rather than a random cation distribution that would normally be implied. \\

\noindent {\textit{Microscopy investigations}}

\begin{figure}
   		 \includegraphics[width=8cm]{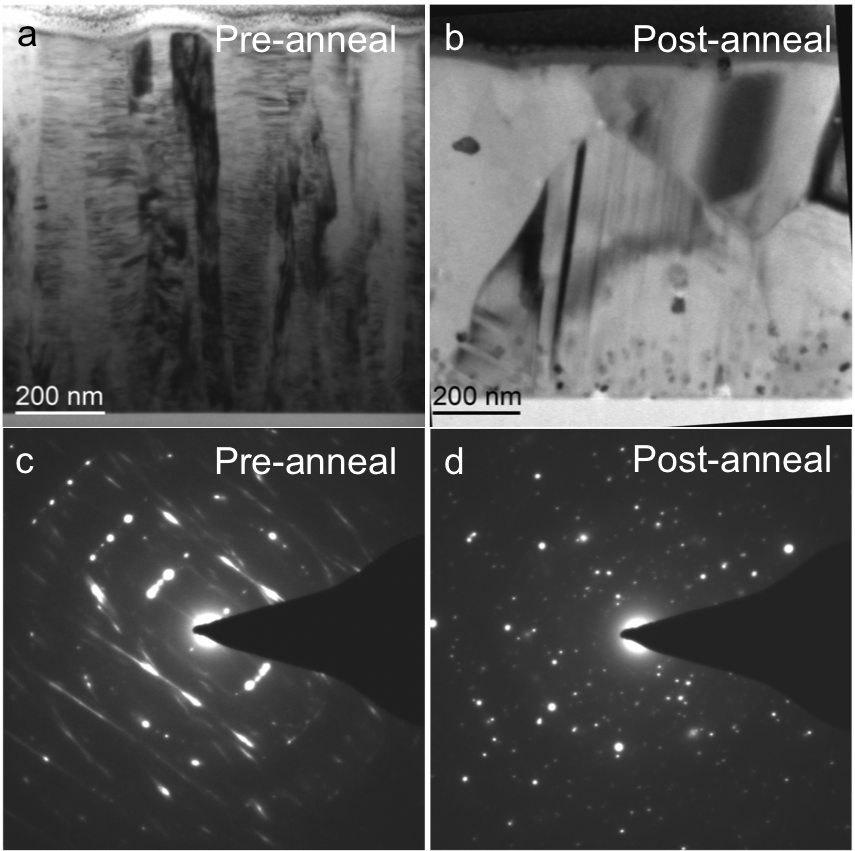} 
\caption{(a) As-deposited films show columnar grains with an extremely high density of \{111\} oriented stacking faults. (b) In the annealed films, the planar defect density is reduced in some grains, but remains high in others. (c) and (d) Tranmission electron diffraction patterns confirm the zinc blende structure of the unannealed sample, and the complex TED pattern of the annealed sample suggests a monoclinic structure.}
\label{Fig:TEM}
\end{figure}

To understand how the monoclinic, ostensibly ``ordered'' sample could still exhibit properties suggesting a disordered structure (such as high $c_1$ values and ps carrier lifetimes), we used transmission electron microscopy (TEM) to examine the microstructure of the two samples. As shown in Fig. \ref{Fig:TEM}a, the as-deposited sample shows columnar grains ($\sim$50 nm across) that span the thickness of the film, and are strongly preferentially oriented in the $<$111$>$ growth direction. All grains have an extremely high density of \{111\} planar defects (stacking faults and twins), which is confirmed by the visible streaks in the transmission electron diffraction (TED) pattern (see Fig. \ref{Fig:TEM}c). The TED patterns also confirm the zinc blende (cubic) crystal structure, as was previously determined from the XRD pattern and Raman spectrum. The annealed sample shows a markedly different grain structure (Fig. \ref{Fig:TEM}b). The grains are larger (200-500 nm) and equiaxed, with no detectable preferential orientation. In this film, some of the grains still have a high density of planar defects, while other grains do not exhibit visible defects.  We note that this sample was annealed for 2 hrs, and the 8 hr anneal resulted in a lower carrier concentration. It is possible that the 8 hr annealed sample has a higher fraction of grains without visible defects, and that even longer annealing times (although impractical for PV absorber fabriction) would further reduce the density of planar defects.  The TED pattern for the annealed sample is complex, suggesting a monoclinic structure (Fig. \ref{Fig:TEM}d).

The effect of stacking faults in Cu$_2$SnS$_3$ can be understood by again considering this structure in terms of S coordination motifs. In the lowest energy state, only two motifs are present: Cu$_2$Sn$_2$-S and Cu$_3$Sn-S. Each motif is coordinated by 12 other motifs, and a coordination number can be used to denote by how many motifs of the same type a certain motif is surrounded. The defect-free monoclinic structure requires that the coordination numbers are set as 2 and 7 for the S-Cu$_2$Sn$_2$ and S-Cu$_3$Sn motifs, respectively. When a stacking fault occurs, it alters the motif coordination numbers, causing local motif clustering and disorder. This is true even in the monoclinic structure; from the TEM, we see that some grains are truly ``ordered'' in that no stacking faults are visible, but other grains still have a high density of stacking faults, and this is ultimately what dominates the carrier transport in these films.

If we now consider the calculated electronic density of states (DOS) for ordered and disordered (i.e., clustered) Cu$_2$SnS$_3$ (Fig. \ref{Fig:IPR}b), the reason for poor minority carrier transport (Fig. \ref{Fig:Thz}) becomes clear. As a result of the high density of stacking faults (Fig. \ref{Fig:TEM}), both samples can be viewed as ``disordered'' when determining electronic properties. The DOS for disordered Cu$_2$SnS$_3$ (Fig. \ref{Fig:IPR}b) shows significant band tailing and a mid-gap state, and both of these could contribute to the picosecond electron decay times observed in the OPTP (Fig. \ref{Fig:Thz}). Additionally, the band tailing reduces the band gap of Cu$_2$SnS$_3$ (0.9-1.35 eV), ultimately reducing the final device performance. Moving forward, control of planar defects and local disorder will be an important research challenge that must be addressed for future improvement of Cu$_2$SnS$_3$ photovoltaic devices. \\

\noindent {\textit{Implications of this work for CZTS and other disordered semiconductors}}

Calcuations of the inverse participation ratio for Cu$_2$ZnSnS$_4$ indicate similarly high degrees of charge localization in the kesterite structure (see Fig. \ref{Fig:CZTS}). The detrimental impacts of disorder in CZTS have been discussed in multiple theoretical works, and include band tailing, potential fluctuations on the order of 200 meV, and the formation of nanoscale compositional inhomogeneities \cite{Zawadzki2015,Gokmen2014,Zawadzki2015_2}. These challenges have proven difficult to address experimentally, and may slow further increases in the efficiency of CZTS-based photovoltaics. Although Cu$_2$SnS$_3$ is not nearly as well studied as CZTS, the similarities between these two materials suggest that Cu$_2$SnS$_3$-based devices may face similar development challenges in the future.

Newer photovoltaic materials with potentially advantageous disorder effects are also being investigated, such as disorder-related band gap tuning in ZnSnN$_2$ \cite{Seryogin1999,Scanlon2012}. These design ideas may offer advantages for the optical properties of the materials; however, the investigations will also need to take into account the effects of disorder on the carrier transport properties. Successful future investigations of disordered semiconductors will require the use of new modeling techniques to accurately assess the electronic effects of disorder, such as those developed in Ref. \cite{Zawadzki2015}. Experimental quantification of the effects of disorder is equally important, using not only structural analysis, but also techniques such as the terahertz spectroscopy and high-resolution TEM used in this study. \\

\noindent {\textbf{Conclusions}} \\

In this study, we investigated the impacts of cation disorder on the electronic and structural properties of Cu$_2$SnS$_3$. We demonstrated the transformation from a cubic to a monoclinic crystal structure upon equilibration of the Cu$_2$SnS$_3$ with SnS, and a concurrent reduction in hole concentration by almost 2 orders of magnitude. We analyzed the as-deposited (cubic/disordered) and annealed (monoclinic/ordered) samples using optical pump terahertz probe spectroscopy, which detected 0.3 and 7 ps decay processes in both samples. By analyzing the complex electrical conductivity from the OPTP measurement, we determined that both samples displayed a high degree of charge localization. When theory was used to calculate the charge localization in ordered and disordered Cu$_2$SnS$_3$, it suggested that charge localization is only found in the disordered structure. TEM investigations revealed high densities of stacking faults and/or twins in both samples, which could be responsible for local disorder and the associated poor minority carrier transport. 

Overall, the results presented in this work identify several challenges to the use of Cu$_2$SnS$_3$ as a photovoltaic absorber. It is possible that different synthesis techniques may result in material without the high density of planar defects exhibited in our sputtered films. However, it is also likely that disorder in Cu$_2$SnS$_3$ is unavoidable due to fundamental thermodynamic reasons (especially when the weak temperature dependence of the disorder is considered \cite{Zawadzki2015}). In this case, further efforts to improve upon the leading 4\% efficient Cu$_2$SnS$_3$ may be hampered by both the existance of local disorder, and the difficulty of detecting this disorder. As research into Cu$_2$SnS$_3$ moves forward, it will be critical to assess the charge localization and disorder effects, and to correlate these properties with photovoltaic device performance.  \\

\noindent {\textbf{Acknowledgements}} \\

This work was supported by the US Department of Energy, Office of Energy Efficiency and Renewable Energy, as a part of the ``Rapid Development of Earth-Abundant Thin Film Solar Cells'' agreement, under Contract No. DE-AC36-08GO28308 to NREL. L.L.B. was supported by the Department of Defense through the National Defense Science and Engineering Graduate Fellowship Program. T.U., R.E., and H.H. gratefully acknowledge the support of this work by the Helmholtz Association Initiative and Network Fund (HNSEI-Project). Thanks to Brenden Ortiz at the Colorado School of Mines for synthesis of SnS and SnS$_2$ powders. Thanks to Lynn Gedvilas and Adam Stokes at the National Renewable Energy Laboratory for Raman spectroscopy and TEM sample prep. Adele Tamboli at the National Renewable Energy Laboratory provided valuable discussion and input.

\clearpage
\noindent {\textbf{Supplementary Information}} \\

\setcounter{figure}{0}

\makeatletter
\renewcommand{\thefigure}{S\@arabic\c@figure}
\makeatother

\setcounter{table}{0}

\makeatletter
\renewcommand{\thetable}{S\@arabic\c@table}
\makeatother

\noindent {\textit{Investigation of annealing conditions}}

We investigated a variety of annealing conditions to determine their effects on the transport properties of Cu$_2$SnS$_3$. Our experimental set up allowed us to easily vary the annealing temperature, atmosphere, and duration. Each of these parameters was investigated independently to determine the effects on the carrier concentration. 

From our previous work, we know that the lowest doping levels in Cu$_2$SnS$_3$ are achieved in Cu-poor and S-poor environments \cite{Baranowski:2014em}. Consulting the Cu-Sn-S chemical potential phase space in Ref. \cite{Zawadzki:2013fr} we see that these conditions are achieved when Cu$_2$SnS$_3$ is in equilibrium with SnS. To verify this theoretical result, we annealed the Cu$_2$SnS$_3$ films under three different atmospheres: Ar only, SnS$_2$+Ar and SnS+Ar (note that SnS sublimes congruently). As expected, the SnS atmosphere resulted in the largest reduction in carrier concentration (see Fig. \ref{Fig:atm}a). The SnS$_2$ atmosphere resulted in moderate carrier concentration reductions, which is likely because the atmosphere was more S-rich than the SnS anneal. Finally, the Ar only atmosphere produced very small reductions in carrier concentration. This can be attributed to the fact that this atmosphere is Sn-poor (analogous to Cu-rich) compared to the other two atmospheres investigated. Concurrent with the decreases in carrier concentration, we observed a slight increase in mobility for the sample annealed in the SnS atmosphere. This increase in mobility is likely connected to a decrease in defect-related scattering caused by the carrier concentration reductions.

To investigate the kinetics of annealing in the SnS atmosphere, we annealed the Cu$_2$SnS$_3$ films at temperatures of 300, 400, and 500$^{\circ}$C, while keeping the anneal duration constant at 30 minutes. Above 500$^{\circ}$C, significant material loss and film delamination was observed. As shown in Fig. \ref{Fig:atm}b, the 400$^{\circ}$C and 500$^{\circ}$C annealing temperatures showed significant reductions in carrier concentration as compared to the 300$^{\circ}$C sample. It is possible that temperatures higher than 500$^{\circ}$C could result in further carrier concentration decreases, but as mentioned above, this was not possible due to material losses at higher temperatures. Thus, we also investigated longer anneal durations at 500$^{\circ}$C. We observed that increasing the anneal duration from 30 minutes to 2 hrs resulted in a lower carrier concentration, and an 8 hr anneal time resulted in further carrier concentration decreases. This indicates that the kinetics of the defect reduction is fairly slow. Although it is possible that further decreases would be observed at anneal times longer than 8 hrs, longer anneals were not investigated in this study.

\begin{figure}
   		 \includegraphics[width=7cm]{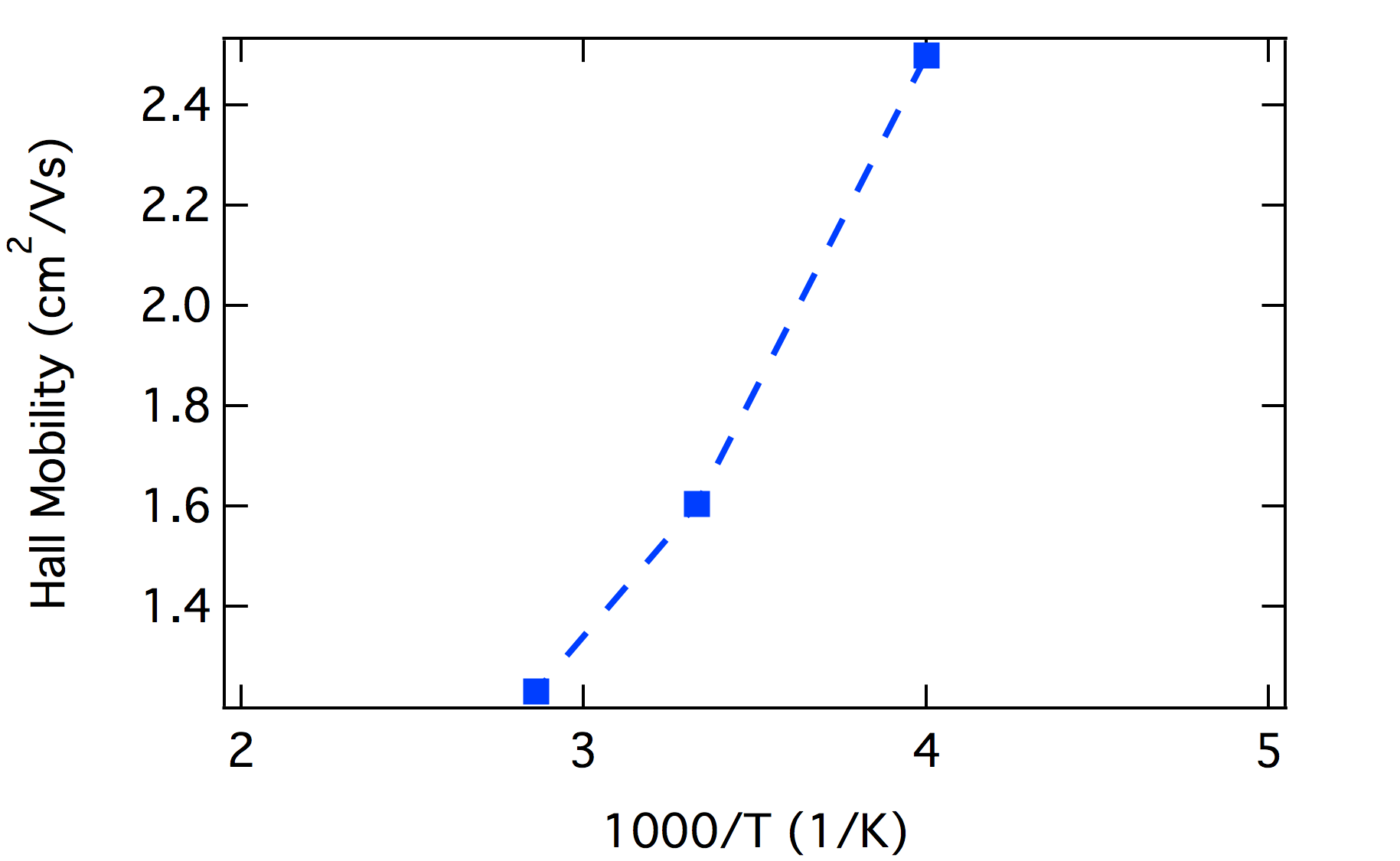}
\caption{The increase in Hall mobility at low temperatures suggests that the hole transport is not grain boundary limited in the pre-annealed samples.}
\label{Fig:tdhall}
\end{figure} 

\clearpage

\begin{figure}
   		 \includegraphics[width=7cm]{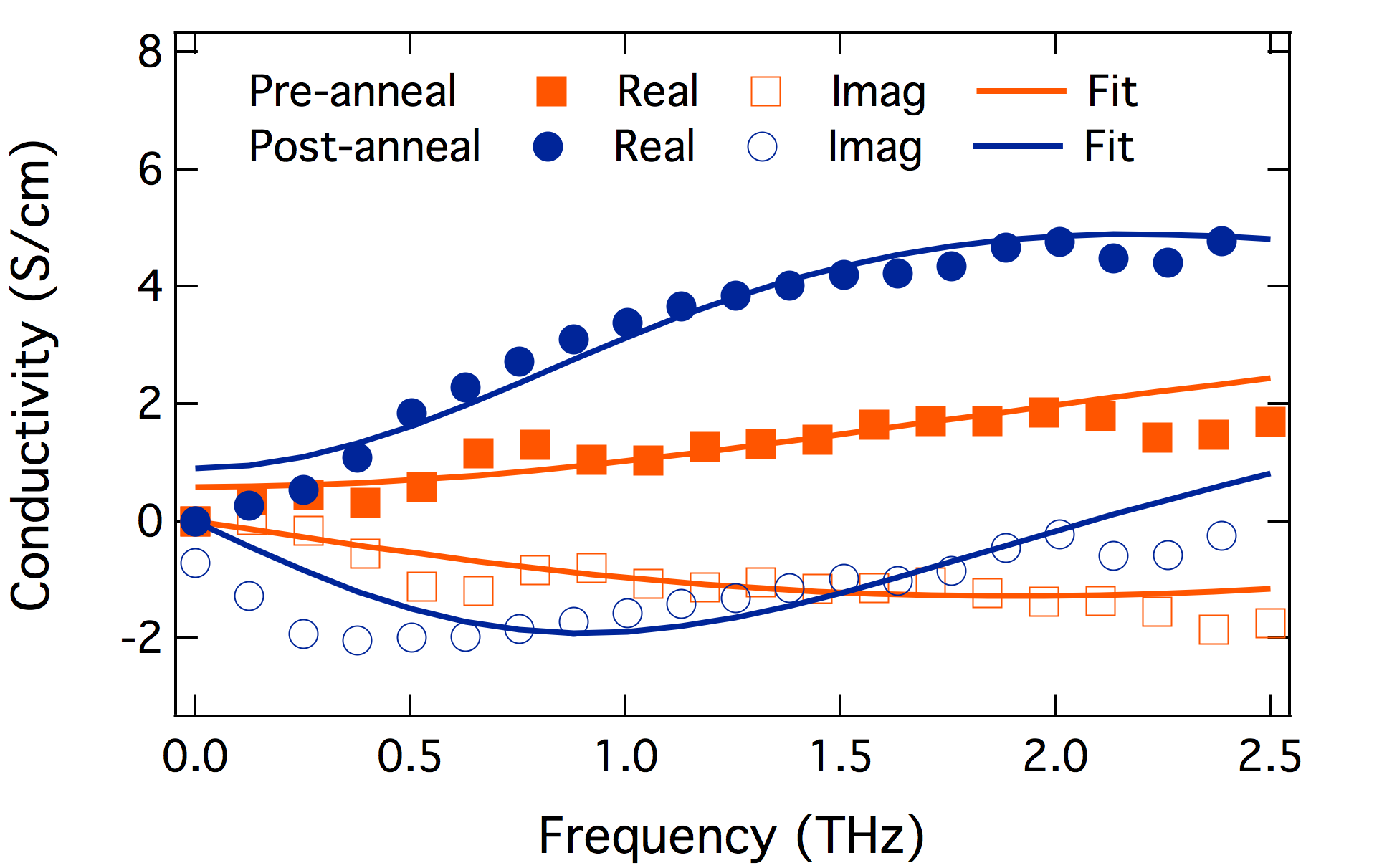} 
\caption{Carrier localization is suggested by the negative imaginary component of the frequency-dependent terahertz conductivity.}
\label{Fig:Thzcond}
\end{figure}

\begin{figure}
   		 \includegraphics[width=7cm]{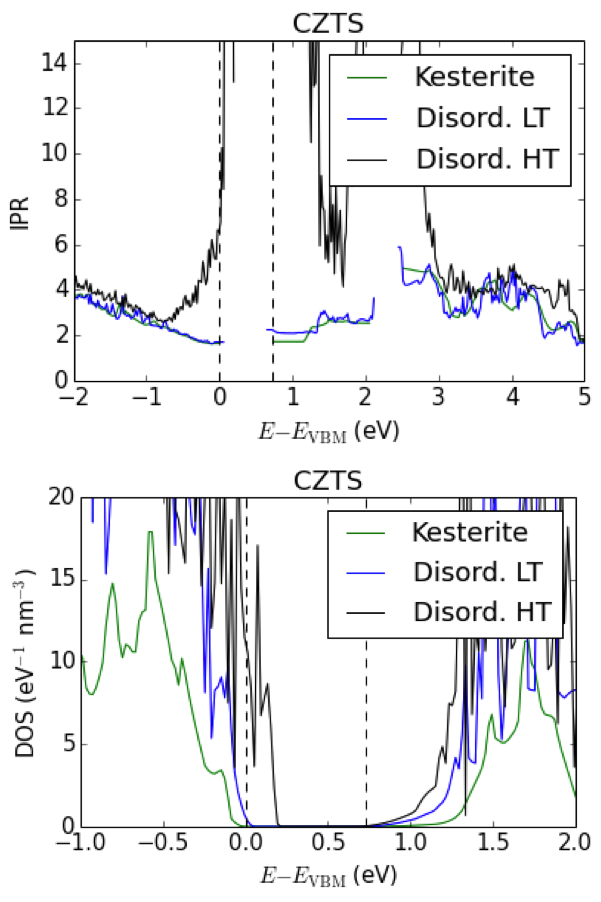} 
\caption{The electronic effects of disorder are even more pronounced in the kesterite structure of CZTS than in Cu$_2$SnS$_3$.}
\label{Fig:CZTS}
\end{figure}

\end{document}